# A Heuristic Approach to the Quantum Measurement Problem: How to Distinguish Particle Detectors from Ordinary Objects

R. Merlin

*Department of Physics, University of Michigan, Ann Arbor, Michigan 48109-1040, USA*

Elementary particle detectors fall broadly into only two classes: phase-transformation devices, such as the bubble chamber, and charge-transfer devices like the Geiger-Müller tube. Quantum measurements are seen to involve transitions from a long-lived metastable state (e. g., superheated liquid or a gas of atoms between charged capacitor plates) to a thermodinamically stable condition. A detector is then a specially prepared object undergoing a metastable-to-stable transformation that is significantly enhanced by the presence of the measured particle, which behaves, in some sense, as the seed of a process of heterogeneous nucleation. Based on this understanding of the operation of a conventional detector, and using results of orthogonality-catastrophe theory, we argue that, in the thermodynamic limit, the pre-measurement Hamiltonian is not the same as that describing the detector during or after the interaction with a particle and, thus, that superpositions of pointer states (Schrödinger's cats) are unphysical because their time evolution is ill defined. Examples of particle-induced changes in the Hamiltonian are also given for ordinary systems whose macroscopic parameters are susceptible to radiation damage, but are not modified by the interaction with a single particle.

Quantum mechanics (QM), in all its variants, provides an extremely accurate and fruitful description of the microscopic world as well as of numerous properties of matter that arise from the collective behavior of its constituents. Its predictions have been confirmed in countless high-precision experiments on physical systems belonging to a wide range of scales, and its development has enabled groundbreaking inventions such as the laser, magnetic resonance imaging and the transistor. In nearly a century, QM has led to the establishment of a vast body of knowledge that accounts for phenomena and concepts as diverse as diamagnetism and superconductivity [1], the quantum-hall effect [2], topological insulators [3], photosynthesis [4], valence and the periodic table, black-body radiation [5], nuclear fission and fusion [6], Bose-Einstein condensation [7], the Higgs mechanism and spontaneous broken symmetry [8], among many others. Yet, in spite of its enormous success, the measurement problem, that is, the question of the interaction of an elementary (or a small composite) particle with a measuring device, still remains as a nagging and unresolved mystery, embodied in the meaning of states such as the Schrödinger's cat, which are seemingly allowed by QM but contradict classical reality [9].

The theory of measurement and, more generally, the theoretical foundations of QM received a huge interest from its founders in the first few decades of the twentieth century [9]. Following a period of reduced attention on fundamental issues, important advances in the understanding of quantum entanglement [10] and macroscopic quantum phenomena [11], coupled with unparalleled progress in the manipulation of atomic-size objects [12] and, more recently, the emergence of quantum information science [13], have brought renewed interest in the subject.

Measurements are described in this paper as scattering processes. Schematic representations of the scattering of an elementary particle (an electron, for definiteness) by an ordinary macroscopic object and a detector are shown, respectively, in Figs. 1(a) and (b). The various possible outcomes of the interaction are represented by Feynman's diagrams which, following the principle of superposition, add up to build the complete solution to Schrödinger's or, more generally, quantum field-theory equations. Although formally identical in regard to scattering, there are crucial differences between ordinary objects and detectors. Broadly speaking, measuring devices are characterized by 'classical' parameters that change as a result of the interaction with a single particle whereas the macroscopic state of ordinary objects is not affected (e. g., a pair of slits coupled to a photon). Standard quantum theory dictates that scattering involving ordinary bodies is deterministic in that different paths can interfere and the process can, in principle, be reversed. On the other hand, and for all practical purposes [14], the scattering by a measuring device leads to the collapse of the particle wavefunction into one of the eigenstates of the particular observable probed by the detector, with a probability given by the modulus square of the corresponding coefficient (Born rule [15]). The measurement problem can thus be formulated as the question of why there are two fundamentally different, deterministic and probabilistic modes in which quantum states can change in time [16].

The Copenhagen and non-local hidden variables [17] interpretation, as well as the many-worlds [18,19], decoherence histories [20,21,22] and environmental decoherence [23,24,25,26] formulations provide distinct answers to the measurement puzzle, as do models based on the dynamical reduction program [27], which invoke modifications of the fundamental equations of the theory. Here, we propose an alternative, heuristic approach based on the fundamental difference between quantum measuring devices and ordinary objects. Relying on a literature survey [28], we argue that there are only two basic classes of detectors for which a measurement conduces either to a phase transformation (Class I) or a macroscopic transfer of charge (Class II). For both types, the state prior to a measurement is quantum mechanically unstable; see below. The role of the particle being measured is to perturb the state vector of the measuring device in such a way that the time it takes for the device to undergo the irreversible transition into a stable state is drastically reduced [29]. More importantly, and relying on arguments borrowed from Anderson's orthogonality-catastrophe theory [30], we contend that the Hamiltonian of the device and, thus, the set of its eigenstates prior to and during (or after) the interaction with the particle are not the same in the thermodynamic limit. More precisely, we argue that predictions of unitary evolution [31] do not hold here because the wavefunction of the whole system cannot be written at all times as a sum of the form

$$\sum_n A_n(q)\Phi_n(\xi) \tag{1}$$

where $\{\Phi_n(\xi)\}$ is a complete set of detector eigenfunctions belonging to the pre-measurement Hilbert space and $A_n(q)$ are functions of the particle variables $q$. Alternatively, what we claim is that the standard dynamics of a quantum state $\Psi$, specified by

$$\left|\Psi(t)\right\rangle = e^{-iHt/\hbar}\left|\Psi(0)\right\rangle , \tag{2}$$

does not apply to the measurement process because the Hamiltonian $H$ is not uniquely defined and, consequently, that Schrödinger's cat-like superpositions involving the initial (or 'ready') and final states of a detector are devoid of physical meaning since they violate the $t$ (time)-evolution rules of QM, as depicted in Eq. (2). Finally, and referring to the diagrams of Fig. 1, these considerations imply that, while non-measuring objects follow the same Schrodinger's equation before, during and after interacting with the particle (however, see later), the deterministic evolution of detectors must terminate at some point, and the wavefunction must collapse, as superpositions involving states associated with different Hamiltonians are forbidden. Although not directly related to the problem of measurement, it is of interest to note that in a transition into a more ordered phase, quantum states associated with different values of the order parameter, that is, different vacua, are described by different energy functionals which, in turn, differ from the Hamiltonian of the disordered phase [32], much as what we claim to be the case of pointer states of a particle detector. Connections between spontaneous broken symmetry and the quantum theory of measurement have been considered in the literature [33,34].

Regardless of the radiation source or device type, a clear definition of what constitutes a measuring system is still lacking. Nevertheless, an assessment of all known detectors [28] clearly indicates that these systems possess extremely long-lived metastable states, which serve as pre-measurement states, and that the particles they probe act as catalysts of the transition into the final stable state, much as kernels do in heterogeneous nucleation [35]. Representative examples of Class I and II particle detectors, namely, the bubble chamber [36] and the Geiger-Müller counter [37], are sketched in Fig. 2(a) and (b), respectively. Bubble chambers are vessels filled with liquid brought into an unstable superheated state just before a measurement. As shown by Mott [38], particles that leave their source as a spherical wave produce straight or, more generally, classical-like tracks of ionized atoms, which serve as condensation centers around which bubbles of the stable phase form. Within the context of Fig. 1b, there is an infinite set of scattering diagrams associated with tracks, and we assert that each track defines a distinct energy-density functional, that is, a different Hamiltonian. Similar considerations apply to the closely-related Wilson cloud chamber and to a photographic plate, which shows phase separation following its development [39]. These as well as track-edge detectors [40] belong to Class I. The Geiger-Müller counter, depicted in Fig. 2(b), is a member of the large group of Class II detectors for which measurements lead ultimately to the transfer of a macroscopic amount of charge from one electrode to another [28]. The transfer is triggered by the particle under observation through,

e. g., the ionization of a few gas atoms (or molecules) or the generation of a few electron-hole pairs in a semiconductor. Photomultipliers, charge-coupled devices, spark and wire chambers, as well as high-energy calorimeters belong to this group. As for the initial state of phase-transformation detectors, the 'ready' state of a Class II device like the Geiger-Müller counter is metastable because there is a large voltage gradient and electrons in the anode have some small but not-zero probability of tunneling across the detector volume and into the cathode (or through the gas, which is weakly conducting). This also applies to the bound electrons in the gas which can undergo field-induced tunneling ionization.

While there are no theories able to account for the whole range of processes by which a single particle leads to a macroscopic modification of a generic device, it is generally agreed that the transfer of energy from the particle to Class I detectors conduces either to a strongly localized trail of damage along the particle path or, specific to bubble (cloud) chambers, to the creation of localized vapor (liquid) nucleation centers, as described by Seitz thermal-spike theory [41]. Similarly, it is also understood that the interaction of particles with Class II devices and, in particular, those relying on charge multiplication, results in the appearance of a localized charge-density fluctuation involving a handful of ionized electron-ion pairs per track, the number of which grows exponentially thanks to the cascade process known as Townsend avalanche [28]. As discussed next, the fact that there are localized constituents in the interactive state is central to our contention, valid in the thermodynamic limit, that this state does not belong to the Hilbert space defined by the pre-interaction Hamiltonian and, therefore, that sums of states involving different pointer values are physically meaningless.

Consider the interaction of a single particle, of mass $M$, with a system of $N \gg 1$ free fermions, of mass $m$, in volume $V$. Although such a system is not strictly a detector because it does not exhibit amplification, it serves to illustrate the concepts delineated in the previous paragraph. The Hamiltonian is $H = H_P + H_F + U$ where $H_P = P^2/2M$ and $H_F = \sum_i p_i^2/2m$ are, respectively, the Hamiltonian of the free particle and that of the fermions while

$$U = \sum_i U(\mathbf{r}_i - \mathbf{R}) \tag{3}$$

describes the interaction. $\mathbf{P}$ ($\mathbf{p}_i$) and $\mathbf{R}$ ($\mathbf{r}_i$) are the momentum and position operators of the particle (*i*th fermion) and $U$ is a finite-range potential. We assume that $M \gg m$ so that the Born-Oppenheimer approximation applies, that is, the eigenstates are approximately of the form $\Phi(\mathbf{r}_i, \mathbf{R})\chi(\mathbf{R})$ where $(H_F + U)\Phi = E(\mathbf{R})\Phi$ and $[H_P + E(\mathbf{R})]\chi = E\chi$. It follows that the single-fermion states before and after the particle enters the region the fermions occupy are, respectively, plane waves and eigenstates of an impurity-like potential. Since, according to orthogonality-catastrophe theory [30], the overlap integral between a many-body state with $U \neq 0$ and an arbitrary Slater-determinant state described solely in terms of plane waves vanishes in the thermodynamic limit $(N \to \infty; N/V = \text{constant})$, quantum superpositions of the form of Eq. (1) are not allowed during the measurement because $U \neq 0$ states are outside the Hilbert space spanned by the eigenvectors of $H_F + H_P$. Identical considerations are likely to apply to the gas of molecules filling a proportional counter or those in a bubble or Wilson chamber. If the pre-measurement state of the gas can be expressed as a sum of extended states, a locally-perturbed state (as that resulting from the interaction with the particle) does not belong to the same Hilbert space. Within this context, and like domains in a phase transition, different localized perturbations (resulting from, e. g., different particle tracks in a bubble chamber) give physically distinct Hamiltonians thereby precluding the existence of the corresponding superpositions. Although our approach does not directly address the question of collapse, we note that adding reduction of the wave-function as a postulate to the theory appears to be the logical and necessary step to solve the problem Hamiltonian multiplicity causes to unitary evolution. By relating collapse to multiplicity, we also circumvent the unsubstantiated practice of invoking collapse in certain situations but not in others.

As for the idealized model of *N*-fermions, Hamiltonian multiplicity also arises in certain actual situations when an elementary particle interacts with an ordinary, non-amplifying object. Consider a defect-free and isotopically-pure crystal of $^{30}$Si that scatters neutrons of an energy such that they can be captured by a silicon atom to eventually become a $^{31}$P donor plus a bound electron [42]. According to the principle of superposition, the combined wavefunction of the crystal after the interaction with a single neutron is of the form $\sum_i B_i(q)|\Upsilon(\mathbf{R}_i)\rangle$, where $|\Upsilon(\mathbf{R}_i)\rangle$ is the state with the phosphorus impurity at lattice site $\mathbf{R}_i$ and $B_i$ are functions of the coordinates $q$ of all the particles involved in the capture. Some reflection shows that such a state is not allowed, however, since each value of $\mathbf{R}_i$ defines a different energy functional, which also differs from that of pure silicon. Moreover, using orthogonality-catastrophe reasons [30] like those considered previously, it is obvious that all the states with one bound electron are orthogonal to all those of an infinite crystal of pure silicon. Analogous considerations apply to a photographic film or plate (an emulsion containing tightly packed crystals of a photosensitive substance) prior to its development. The photon detection process in, say, AgBr plates, begins with the promotion of an electron to a conduction-band state of one of the crystallites from where it first falls into a shallow trap and then combines with an interstitial $Ag^+$ ion to produce a neutral silver atom [43]. An undeveloped film, prior amplification, is not a measuring device. If more than one photon is captured so that a speck forms containing more than 2-3 atoms of neutral silver, it becomes a so-called latent image, which can then undergo development to turn the entire crystallite into metal-lic silver [39]. As for the silicon example, neutral-silver states associated with different crystallites belong to different Hamiltonians so that superpositions where a single neutral atom is in various crystallites at the same time are physically unsound.

To summarize, we have proposed a novel, heuristic approach to solving the quantum measurement problem, which applies to isolated as well as to systems in contact with the environment. Our main assertion is that there is a diversity of energy functionals, i. e., Hamiltonians, associated with different pointer states of a detector, thereby establishing a relationship between particle tracks to domains in a symmetry-broken phase. Hamiltonian multiplicity both precludes the realization of quantum states of the Schrödinger-cat type and requires the introduction of wavefunction collapse to resolve the problem it imposes on unitary evolution. We have identified metastability of the pre-measurement state and amplification as the key characteristics of a detector and, by relating quantum measurements to heterogeneous nucleation, clarified the role of the particle as the trigger that strongly increases the transition rate into the stable state. Since processes such as charge separation, super-heating and cooling are naturally occurring phenomena, human observers are thus removed from the measurement process. Finally, by using the examples of the photographic plate and the capture of a neutron by a piece of silicon, we have shown that amplification is not a necessary condition for the reduction of the wavefunction as the interaction of a single particle with an ordinary object can also lead to forbidden superpositions.

# FIGURE CAPTIONS

**FIG. 1** (color online). Scattering-process representations of the interaction between (a) an ordinary object and (b) a particle detector with a single electron. A few scattering channels described by Feynman diagrams are shown. Ordinary objects are represented by a mirror and a receptacle containing a substance whose refractive index is large enough for the electron to emit Cherenkov radiation as it traverses the medium. Double-line arrows denote quantum states of either the ordinary object (OO) or the measuring device (MD).

**FIG. 2** (color online). Schematic diagrams representing the before- and after-the-measurement states of (a) a bubble chamber and (b) a Geiger-Müller counter.

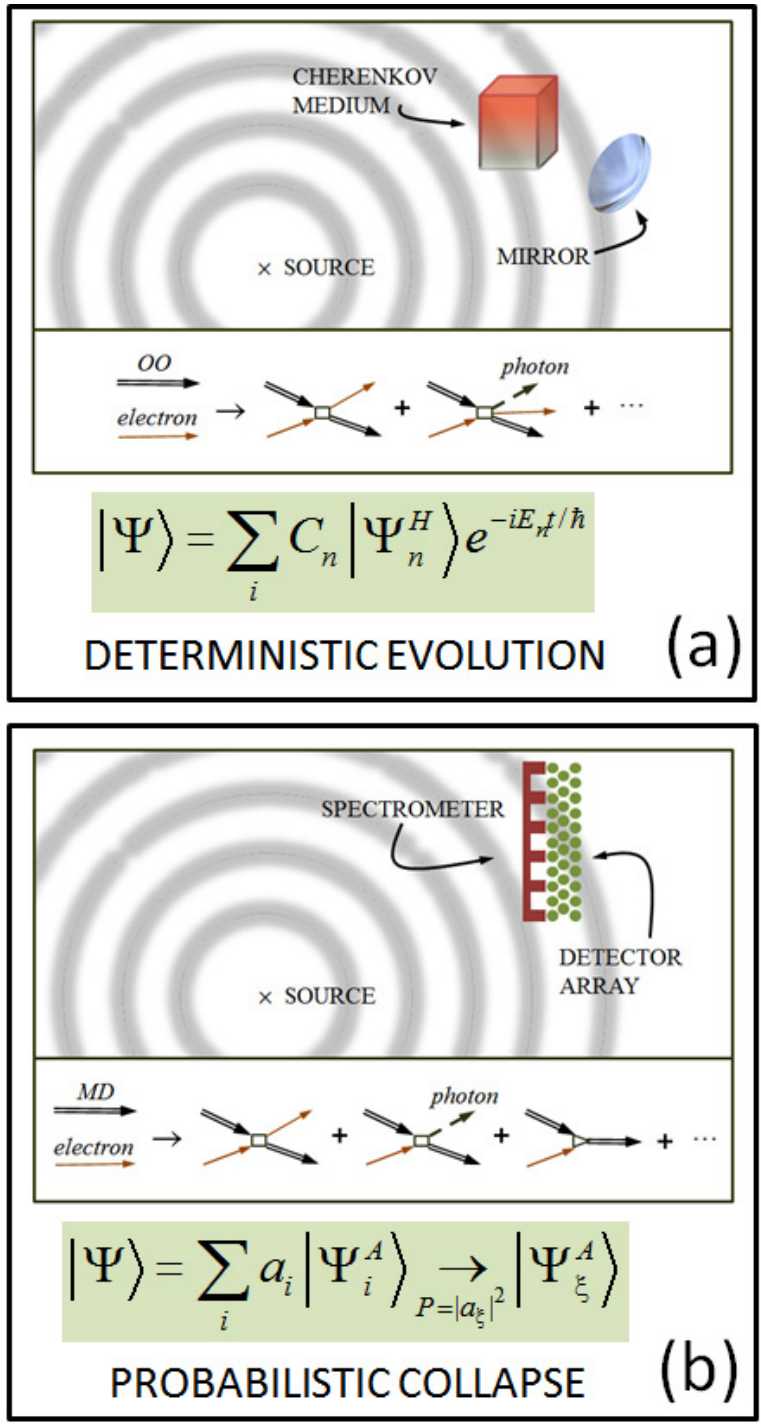

Figure 1

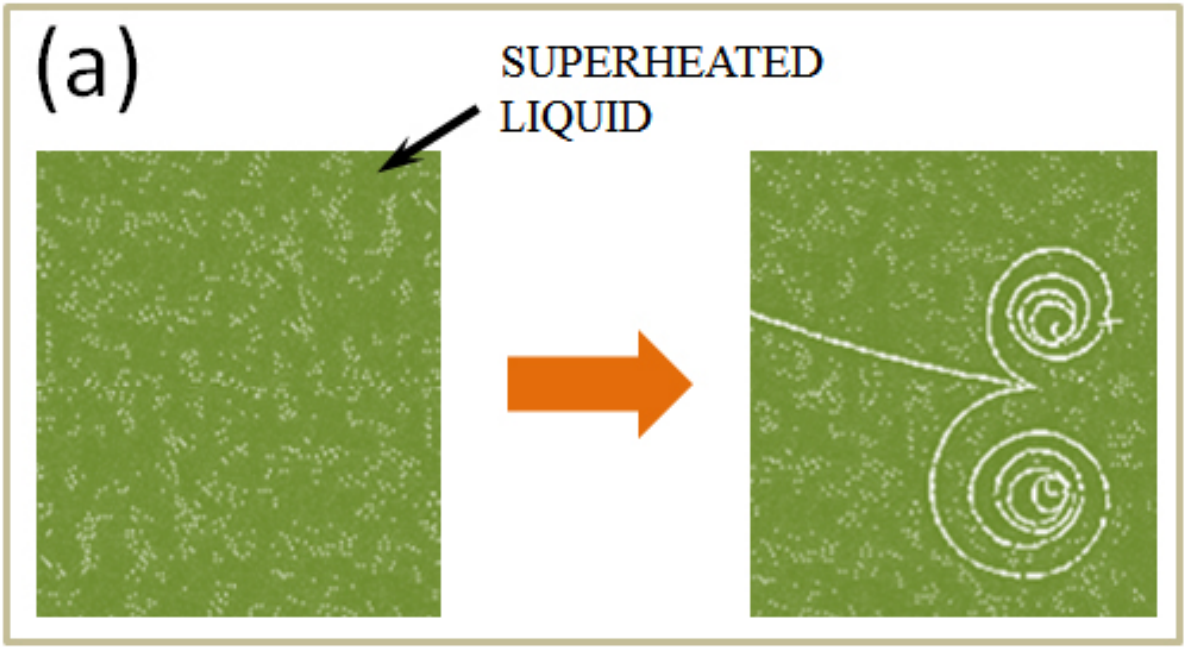

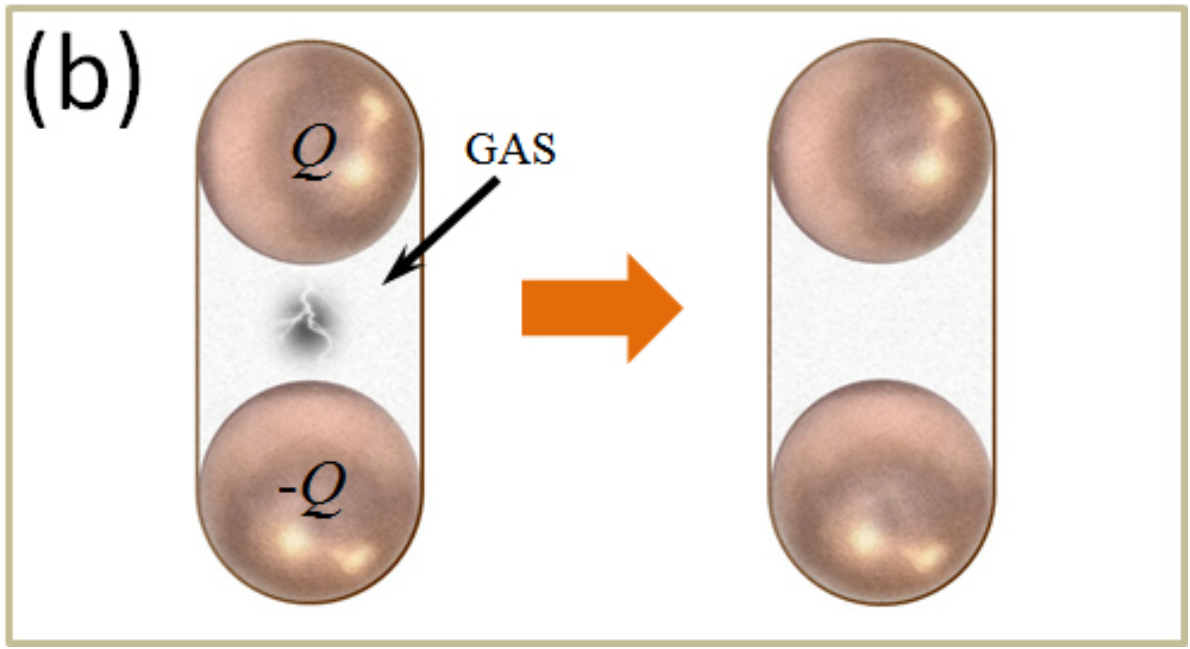

Figure 2